\documentclass[11pt]{article} 
\usepackage{moriond,graphicx} 

\def\be{\begin{equation}} 
\def\ee{\end{equation}} 
\def\bea{\begin{eqnarray}} 
\def\eea{\end{eqnarray}} 
 
\begin{document} 
\vspace*{4cm} 
\title{SUPERNOVA NEUTRINOS, LSND AND MINIBOONE} 
 
\author{ MICHEL SOREL } 
 
\address{Department of Physics, Columbia University, New York,
 NY 10027, USA} 
 
\maketitle\abstracts{
The unique constraints on neutrino oscillations
 which can be obtained by measuring the energy spectrum of supernova $\bar{\nu}_e$'s are
 first discussed. The focus is on 4-neutrino mass and mixing models capable of explaining
 the LSND evidence of $\bar{\nu}_{\mu}\rightarrow \bar{\nu}_e$ oscillations.
 The potential of the Fermilab neutrino detector MiniBooNE to observe supernova
 neutrinos is then addressed, and a general description of the detector
 is given. As of May, 2002, the MiniBooNE detector is fully operational.}

\section{Introduction}
Supernovae (SNe) are extremely violent and luminous stellar explosions in which the optical
 luminosity of the star at maximum light emission can be as great as that of a small
 galaxy. There are two entirely different classes of SN explosions. SNe of type Ia
 are the cosmologically interesting ones,
 used as standard candles by cosmologists in their quest to understand
 the space-time structure of the Universe. SNe Ia are the most luminous optically,
 but do not have a significant neutrino emission.
 This presentation focuses exclusively on the other class of SNe, comprising the
 types II, Ib and Ic, which are known as ``core-collapse'' SNe. \cite{raffelt} Core-collapse SNe
 produce an intense and detectable neutrino burst emission. \\
\indent In a core-collapse SN explosion, neutrinos and antineutrinos of all flavors
 are produced via: $NN\rightarrow NN\nu\bar{\nu}$, $e^+e^-\rightarrow \nu\bar{\nu}$,
 and other processes. Because of the very high matter density at the
 neutrino production site, neutrinos get trapped for some time and reach thermal equilibrium
 with the surrounding material. Neutrinos eventually escape over a typical timescale of
 $1-10$ seconds and, ignoring neutrino oscillations,
 each flavor of neutrinos or antineutrinos would take away approximately
 the same fraction of energy, that is $1/6$ of the total. However, the average
 neutrino energies at the neutrinospheres, {\it i.e.} the surfaces of
 last scattering for the neutrinos, differ from one flavor to another. The
 average energies are estimated to be $10-13$MeV for $\nu_e$, $14-17$MeV for
 $\bar{\nu}_e$, $23-27$MeV for $\nu_{\mu ,\tau},\bar{\nu}_{\mu ,\tau}$.
 \cite{raffelt,janka} \\
\indent There are several important reasons to try to detect neutrinos from
 core-collapse SNe on Earth. First, high-energy astrophysicists can better understand
 the complex core-collapse explosion mechanism itself, by looking at
 its neutrino emission. This is because about $99\%$ of the gravitational binding energy
 liberated in a SN explosion, $\sim 3\cdot 10^{53}\mbox{ erg}$, is emitted
 under the form of neutrinos. Second, given the much smaller neutrino
 cross-sections in the SN mantle compared to the photon cross-sections,
 neutrinos from a SN explosion reach Earth on the order of a few hours before
 the corresponding optical signature. An early warning could then be sent to
 astronomers, and the early stages of the optical SN explosion be observed.
 Finally, SN neutrinos provide an additional (and, in some respects, unique)
 handle for particle-physicists to discern neutrino masses and mixing. \\
\indent The paper is organized as follows. In Section 2, the observation
 of SN $\bar{\nu}_e$ energy spectra as a probe for neutrino oscillations
 is discussed. The focus is on 4-neutrino mass and mixing models explaining
 the LSND evidence for $\bar{\nu}_{\mu}\rightarrow \nu_e$ oscillations.
 See Ref.\cite{sorel} for more details. Section 3 describes the
 MiniBooNE detector located at Fermilab, discussing in some detail
 its SN neutrino detection potential. Ref.\cite{sharp} gives more details
 on MiniBooNE as a SN neutrino detector.
%
%
\section{Supernova neutrino energy spectra as a probe for neutrino
 oscillations}
 The SN models predictions for the neutrino energy spectra at production can be confronted with
 the observation of the SN $\bar{\nu}_e$
 energy spectrum detected on Earth. In this section, the focus is on antineutrino oscillations
 (as opposed to neutrino oscillations) since the $\bar{\nu}_e$ detection on Earth is the easiest,
 as demonstrated for example for MiniBooNE in Section 3. Neutrino oscillations are expected to modify
 the spectrum since
 $\langle E_{\bar{\nu}_e}\rangle < \langle E_{\bar{\nu}_{\mu},\bar{\nu}_{\tau}}\rangle$.
 The energy-dependence of the neutrino cross-section in the detector material \cite{vogel},
 approximately $\sigma_{\bar{\nu}_ep}\propto (E_{\bar{\nu}_e}-1.29\mbox{MeV})^2$, helps
 in making the $\bar{\nu}_e$ energy spectrum distortion a sensitive experimental probe
 to neutrino oscillations. Quite interestingly, the extent of the spectrum modification depends
 crucially on the specifics of the neutrino mixing scheme and on the neutrino mass hierarchy under
 consideration.
\subsection{4-neutrino models to explain neutrino oscillations}
As it is well-known, we now have three experimental hints pointing toward neutrino
 oscillations: solar, atmospheric, and LSND oscillations. While the first two signatures
 are considered to be well-established, the LSND signal \cite{aguilar} needs verification and should be
 definitely confirmed or refuted by the MiniBooNE experiment at Fermilab (see Section 3 and
 Ref.\cite{bazarko}).
 Assuming CPT-invariance, in order to explain all of the data via oscillations, (at least) four
 neutrinos are needed. In these 4-neutrino models, a {\it sterile} neutrino $\bar{\nu}_s$
 with no standard weak couplings is added to the three {\it active} neutrino flavors
 $\bar{\nu}_e,\bar{\nu}_{\mu} ,\bar{\nu}_{\tau}$. Apart from MiniBooNE, a ``neutrino factory''
 as currently foreseen should
 also be able to distinguish $4$-neutrino models from $3$-neutrino ones, that is neutrino models
 without a sterile state.\cite{meloni} \\    
\begin{figure}[tb]
\includegraphics[width=3.9cm]{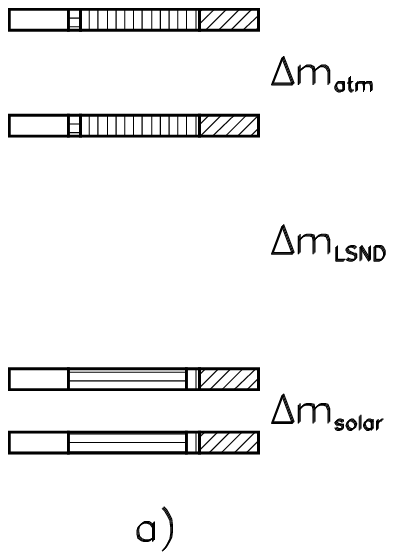} 
\includegraphics[width=3.9cm]{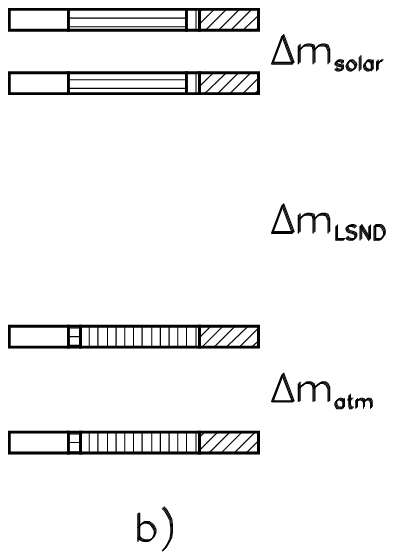} 
\includegraphics[width=3.9cm]{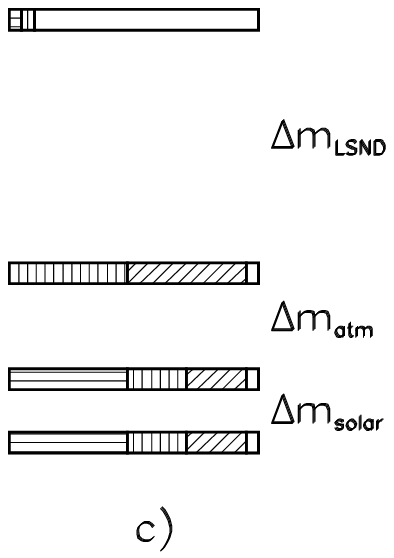} 
\includegraphics[width=3.9cm]{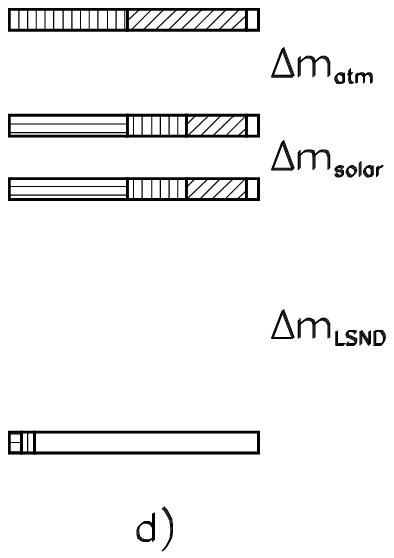}
\caption{\label{fig:numodels}Flavor content of the four neutrino mass mass eigenstates for four
 $4$-neutrino models. Masses increase from bottom to top. Empty rectangles: $\bar{\nu}_s$,
 horizontally-hatched: $\bar{\nu}_e$, vertically-hatched: $\bar{\nu}_{\mu}$,
 diagonally-hatched: $\bar{\nu}_{\tau}$. a) normal $(2+2)$ model; b) LSND-inverted $(2+2)$;
 c) normal $(3+1)$; d) LSND-inverted $(3+1)$. See text for the definition of ``LSND-inverted'' hierarchy.}
\end{figure} 
\indent 4-neutrino models fall into two categories: either $(2+2)$, or $(3+1)$ models.
 The flavor content of the four neutrino mass eigenstates involved in these models are
 shown in Fig.\ref{fig:numodels}. In $(2+2)$ models, Figs.\ref{fig:numodels}a and b,
 the doublet of mass eigenstates
 responsible for solar oscillations and the doublet responsible for atmospheric
 neutrino oscillations are widely separated by the LSND mass gap. In $(3+1)$ models,
 Figs.\ref{fig:numodels}c and d,
 the triplet of mas seigenstates responsible for solar and atmospheric oscillations
 and an almost entirely sterile state are separated by the LSND mass gap. Moreover, given
 the present experimental input on neutrino oscillations from shot-baseline experiments,
 we have no information about the sign of the mass difference $\Delta m_{LSND}$
 explaining LSND oscillations. Therefore, in addition to the ``normal'' hierarchies
 depicted in Figs.\ref{fig:numodels}a and c, we consider also the ``LSND-inverted''
 hierarchies of Figs.\ref{fig:numodels}b and d, obtained from the normal ones by
 substituting $\Delta m_{LSND}\rightarrow \Delta m_{LSND}$.
\subsection{Matter effects and oscillations in the supernova environment}
In matter, $\bar{\nu}_e$'s undergo coherent CC forward-scattering from electrons, while
 all active flavors $\bar{\nu}$'s undergo coherent NC forward-scattering from electrons,
 protons, and neutrons. These processes give rise to an interaction potential
 $V=V_W+V_Z$, which has to be added to the Hamiltonian $H_0$ describing neutrino
 propagation in vacuum. In the flavor basis, we can write the Hamiltonian describing
 neutrino propagation in matter as:\cite{kayser}
\begin{equation}
(H)_{\alpha\beta}=(H_0+V)_{\alpha\beta} = U^*_{\alpha i}U^*_{\beta i}\frac{m_i^2}{2p}+
 A_{\alpha}\frac{G_F \rho}{m_N}\delta_{\alpha\beta}
\label{eqn:matter}
\end{equation}
\noindent where $U$ is the mixing matrix relating the neutrino mass eigenstates $\bar{\nu}_i$
 to the flavor eigenstates $\bar{\nu}_{\alpha},\ \alpha =e,\mu ,\tau ,s$, via
 $|\bar{\nu}_{\alpha}\rangle = U_{\alpha i}|\bar{\nu}_i\rangle$, $m_i$ is the mass of the
 neutrino mass eigenstate $\bar{\nu}_i$, $p$ is the neutrino momentum, $A_{\alpha}$ is
 a flavor-dependent numerical coefficient which can be calculated \cite{sorel} and gives
 $A_{\mu}>A_{\tau}>A_s>A_e$, $G_F$ is the Fermi constant, $\rho$ is the matter density,
 and $m_N$ is the nucleon mass.
 In Eq.\ref{eqn:matter}, we have neglected a term proportional to the identity matrix,
 since it is irrelevant for neutrino oscillations. \\
\indent At the neutrinosphere, the matter density is so high ($\rho \sim 10^{12}g/cm^3$)
 that the interaction potential $V$ dominates over the vacuum Hamiltonian $H_0$, so that the
 propagation eigenstates coincide with the flavor eigenstates. As the propagation eigenstates
 free-stream outwards, toward regions of lower density in the SN environment, their flavor
 composition changes, ultimately reaching the flavor composition of the mass eigenstates in the
 vacuum. Given that neutrinos the neutrinos escape the SN as mass eigenstates, no further
 flavor oscillations occur on their path to the Earth.
 Therefore, flavor oscillations can occur: the probability $p_{\alpha\rightarrow \beta}$
 for a state born as a $\bar{\nu}_{\alpha}$ to reach Earth behaving as a $\bar{\nu}_{\beta}$
 is nonzero for $\alpha \neq \beta$. These probabilities $p_{\alpha\rightarrow \beta}$ can
 be calculated \cite{sorel} for all flavor combinations ($\alpha ,\beta =e,\mu ,\tau ,s$), given   
 a certain neutrino mixing matrix and hierarchy, and using
 the {\it adiabatic approximation}. Antineutrinos propagate adiabatically
 if the varying matter density they encounter changes slowly enough so that transitions between
 local (instantaneous) Hamiltonian eigenstates can be neglected throughout the entire
 antineutrino propagation. From these probabilities,
 one finds that the inverted schemes in Figs.\ref{fig:numodels}b and d
 predict large $\bar{\nu}_{\mu ,\tau}\rightarrow \bar{\nu}_e$ conversions, and therefore
 large distortions to the $\bar{\nu}_e$ energy spectrum observed on Earth. On the contrary,
 the normal schemes in Figs.\ref{fig:numodels}a and c predict little distortions to the
 $\bar{\nu}_e$ spectrum caused by oscillations.
\subsection{Application: constraints on oscillations from SN1987A}
On Feb 23rd, 1987, SN1987A exploded in the Large Magellanic Cloud, a small satellite
 galaxy of the Milky Way. Its neutrino emission was detected by at least two neutrino
 detectors: Kamiokande-2 in Japan, and IMB-3 in the US. Overall, twenty neutrino interactions
 were detected, all of which consistent with $\bar{\nu}_e$ interactions.\cite{koshiba}
 From the measured
 energy spectrum, a low-energy flux $F_{\bar{\nu}_e}$ was inferred, consistent with
 no oscillations. If we define the primary fluxes at production for $\bar{\nu}_{\mu}$
 (or, equivalently, $\bar{\nu}_{\tau}$) and $\bar{\nu}_e$ as $F^0_{\bar{\nu}_{\mu}}$
 and $F^0_{\bar{\nu}_{\tau}}$ respectively, and the {\it permutation factor} $p$
 as the high-energy admixture in the flux on Earth $F_{\bar{\nu}_e}$ due to
 $\bar{\nu}_{\mu ,\tau}\rightarrow \bar{\nu}_e$ oscillations:
\begin{equation}
F_{\bar{\nu}_e}(E) \propto (pF^0_{\bar{\nu}_{\mu}}(E)+(1-p)F^0_{\bar{\nu}_e}(E)
\end{equation}
\noindent a conservative upper limit of $p<0.22$ at $99\%$ CL can be assumed.\cite{sorel} \\
\indent In conclusion, based on SN1987A neutrino data only, the inverted hierarchies of
 Figs.\ref{fig:numodels}b and d are excluded. On the other hand, for the normal mass hierarchy
 schemes in Figs.\ref{fig:numodels}a and c, SN1987A data do not provide competitive
 constraints with respect to existing constraints from reactor, accelerator, atmospheric
 and solar neutrinos; additional experimental input is necessary to unambiguously discern
 the neutrino mass and mixing properties. Undoubtedly, the detection of supernova neutrinos by present
 or near-term experiments such as MiniBooNE would prove very useful in this respect.
%
%
\section{The MiniBooNE detector and its potential for detecting supernova neutrinos}
\subsection{Main motivation of the MiniBooNE experiment}
The MiniBooNE experiment located at Fermilab is designed to be a definitive
 investigation of
the LSND evidence for $\bar{\nu}_{\mu}\rightarrow \bar{\nu}_e$
oscillations, which is the first accelerator-based evidence for
oscillations. The detector consists of a $12$m diameter spherical tank
covered on the inside by 1280 $20$cm diameter phototubes in the detector
region and by 240 phototubes in the $>99\%$ efficient veto region.
 The tank is filled with
$0.8$ktons of pure mineral oil, giving a fiducial volume mass of $0.44$ktons.
The detector is located $500$m downstream of a new neutrino source that
is fed by the $8$GeV proton Booster. A $50$m decay pipe following the Be
target and magnetic focusing horn allows secondary pions to decay into
$\nu_{\mu}$'s with an average energy of about $1$GeV. By switching the horn
polarity, a beam that is predominantely $\bar{\nu}_{\mu}$'s can be produced.
An intermediate absorber can be moved into and out of the beam at a distance
of $25$m, to allow a systematic check of the backgrounds.
With $5\cdot10^{20}$ protons on target (about a year at
design intensity), MiniBooNE will be able to cover the entire LSND allowed
region with high sensitivity ($>5\sigma$). As of May, 2002, the detector is
 fully operational. High-intensity beam data-taking will start in July, 2002.
\subsection{MiniBooNE as a supernova neutrino detector}
\begin{figure}[tb]
\includegraphics[width=7.8cm]{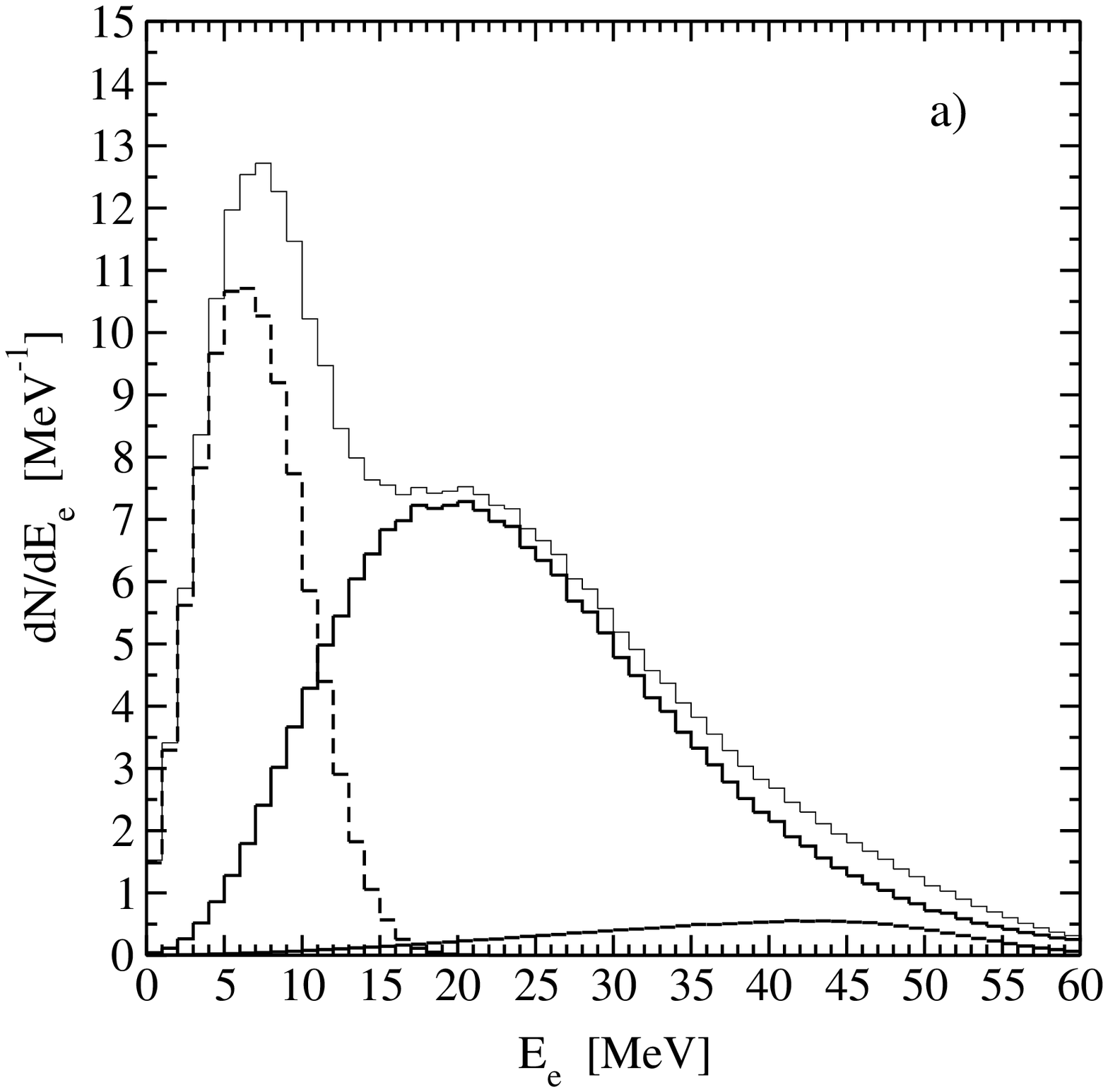} \hspace{0.5cm}
\includegraphics[width=7.8cm]{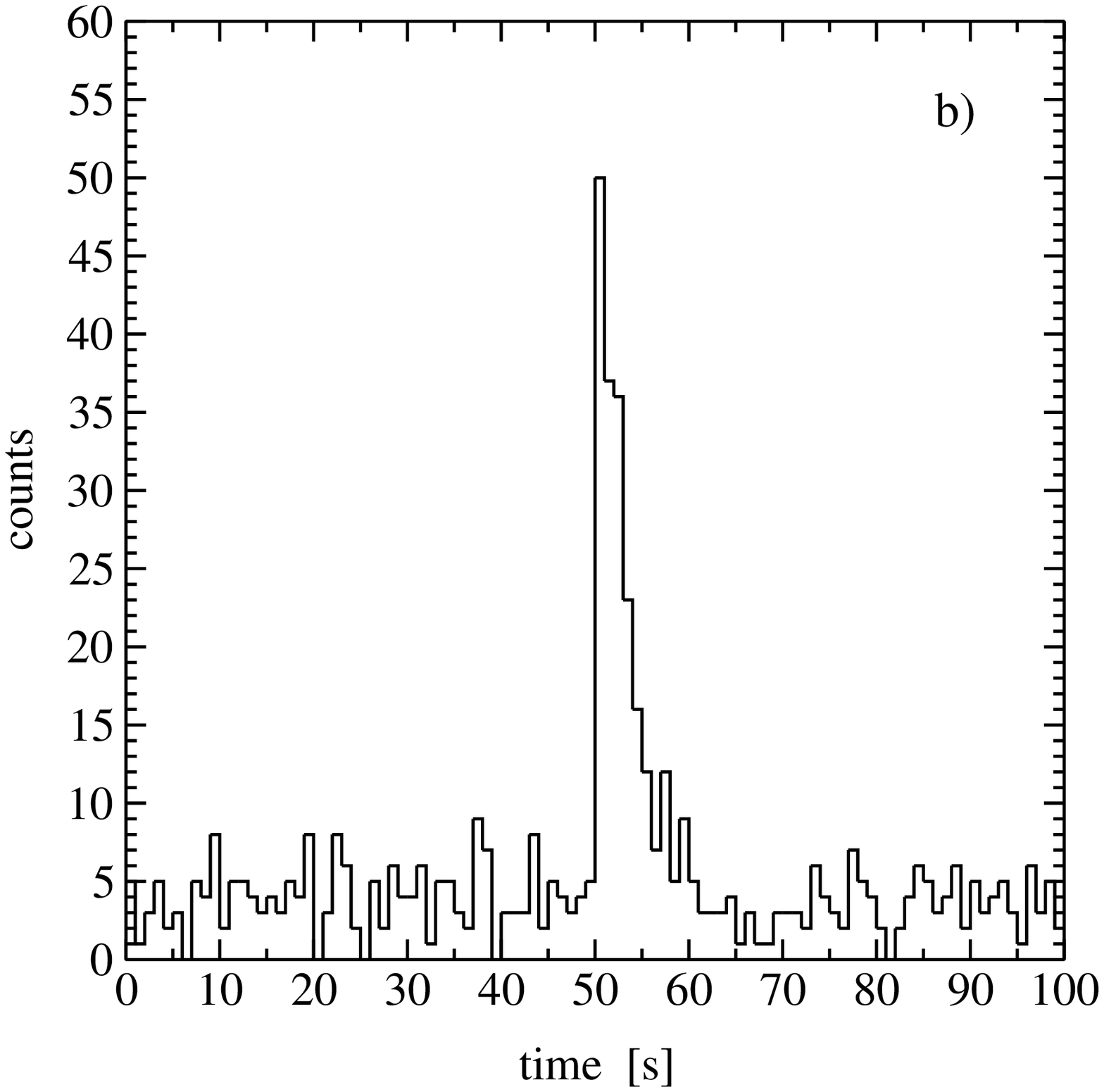}
\caption{\label{fig:boone} a) Spectra of the supernova signal (solid histogram peaking
 at $20$MeV), $^{12}B$ decay background (dashed
 histogram peaking at low energy), and surviving muon decay background (solid histogram peaking at about $50$MeV)
 in MiniBooNE versus the true positron total energy $E_e$, over a $10$s interval
 assumed to contain the supernova signal. The thin solid histogram indicates the spectrum summed over signal and
 backgrounds. Detector deadtime and energy resolution are taken into account.
 Below about $5$MeV, backgrounds from natural radioactivity
 will dominate over the spectra shown. b) Total number of signal and background events, in $1$s bins over
 $100$s, showing the Poisson fluctuations for one random simulated experiment with a
 supernova at $t=50$s. The background
 rate shown is after the cuts described in the text. The supernova signal was modeled as a sharp
 rise followed by an exponential decay with time constant of $3$s.}
\end{figure}

This section is heavily borrowed from Ref.\cite{sharp}.
MiniBooNE can also be used as a SN neutrino detector without interfering with the
 main accelerator-based oscillation search. This is because the Booster beam will
 operate with a duty factor of $8\cdot 10^{-6}$; for the rest of the time,
 MiniBooNE will be functioning as a SN neutrino detector.
 Given the requirements for the accelerator-based search, the
 MiniBooNE detector was built with a dirt overburden of only $3$m. This
 is enough to nearly
 eliminate the hadronic component in cosmic rays, and setting the total
 cosmic-ray muon rate in the detector at the level of $10$kHz, with about
 $8$kHz throughgoing and $2$kHz stopping. Despite the likely skepticism,
 it has been shown a surface-level detector such as MiniBooNE can reduce
 its cosmic-ray muon backgrounds enough to function as a SN neutrino detector.\cite{sharp} \\
\indent The SN neutrino detection reaction in MiniBooNE is $\bar{\nu}_e+p\rightarrow e^++n$.
 The positrons will be detected in MiniBooNE by their Cherenkov (and scintillation) light.
 The neutrons will be radiatively captured on protons, but the resulting $2.2$MeV gamma
 rays will not be visible, due to low-energy radioactivity backgrounds. The expected \cite{sharp}
 number $N$ of events in MiniBooNE from a galactic SN explosion, at a distance $D=10$kpc and
 releasing a gravitational binding energy $E_B=3\cdot 10^{53}$erg, is $N\simeq 230$. This number
 assumes that all events within a radius of $5.5$m from the center of the detector can be used,
 corresponding to
 a detector mass of $0.595$ktons. The next-most important reaction in the detector will be the NC nuclear
 excitation of $^{12}C$, which yields a $15.11$MeV gamma, with only $N\simeq 30$ events expected
 \cite{sharp} and probably beyond reach. \\
\indent In addition to detect as many SN neutrinos as possible, it is also desirable
 to measure the energy of the corresponding positrons accurately. Neglecting nuclear recoil,
 the relation between positron energy and neutrino energy is very simple:
 $E_{e^+}=E_{\bar{\nu}_e}-1.29$MeV. One of the motivations for measuring the positron (and, therefore,
 neutrino) energy is given
 in Section 2, that is to constrain neutrino oscillations. The relative energy
 resolution in MiniBooNE
 should be \cite{sharp} $\delta /E \simeq 0.75/\sqrt{E}$, where all energies are in MeV, and therefore
 reasonably good even at low, $\sim 20$MeV, SN energies. \\
\indent The key question, of course, is whether these signal events can be separated from the
 cosmic-ray related backgrounds expected in MiniBooNE. The two main backgrounds in the
 energy window of interest for SN neutrino detection are due to cosmic-ray
 muons decaying at rest, and to cosmic-ray muons capturing on $^{12}C$ nuclei followed by
 the beta decay of the resulting $^{12}B$ nuclei.
The first background is due to Michel electrons, and can be reduced dramatically by imposing
 a holdoff of $15.2\mu s$ after every muon (the muon lifetime is $2.2\mu s$). With this holdoff,
 all but a $10^{-3}$ fraction of Michel decays are cut, so that the true rate of surviving Michels
 will be a small $2$Hz in the main detector volume. The holdoff introduces a manageable deadtime
 fraction of $0.15$.
The second background is due to $\beta^-$-decay of $^{12}B$ nuclei. This decay has
 a very long lifetime of $20$ms, and therefore a holdoff cannot be used. The rate for
 this background, integrated over all electron energies, is about $11$Hz. However, most
 of the $^{12}B$-decay electrons will have energies well below the typical SN event energies.
 Background rates from radioactive sources are negligible by cutting $<5$MeV events.
 The correctly normalized energy spectra for a galactic SN signal and for the two main backgrounds
 considered are shown in the Fig.\ref{fig:boone}a taken from Ref.\cite{sharp},
 where the energy resolution of the MiniBooNE detector is also taken into account. \\
\indent In sum, by cutting events below $10$MeV, all of the backgrounds due to natural radioactivity
 and $\simeq 80\%$
 of the $^{12}B$ decay background can be cut. The steady-state, background, rate should be about $4$Hz.
 A candidate SN can be flagged by a large increase in the data rate, as shown in the Fig.\ref{fig:boone}b,
 also taken from Ref.\cite{sharp}. \\
\indent What can MiniBooNE add to the woldwide effort \cite{others} to detect SN neutrinos?
 First, it is highly
 desirable to have as many different detectors as possible. This will allow important cross checks of
 the results. Second, MiniBooNE may be able to act as a node in the SuperNova Early Warning System
 (SNEWS). Having many independent nodes in the network greatly reduces the false alarm rate.
 Third, not all detectors are live all the time, due to upgrades and calibrations. Until Super-Kamiokande
 is being repaired, the $\bar{\nu}_e+p\rightarrow e^++n$ in MiniBooNE would be comparable to that
 from other detectors with hydrogen targets.
%
%
\section{Summary}
The phenomenology and observation of supernova neutrinos is an interdisciplinary topic, of interest to
 astronomers, high-energy astrophysicsts, particle physicists. From the particle physics point of view,
 the observation of supernova neutrinos can provide unique constraints on the neutrino mass and mixing
 schemes explaining neutrino oscillations. The relation between the
 measured $\bar{\nu}_e$ energy spectra and various $4$-neutrino models explaining the solar,
 atmospheric and LSND signatures for neutrino oscillations was discussed.\cite{sorel} \\
\indent Moreover, the supernova neutrino detection capabilities of the MiniBooNE detector were
 specifically addressed.\cite{sharp} It was shown that, despite being optimized for higher energies and
 being at a shallow depth of only $3$m, the MiniBooNE detector can in principle clearly separate
 the neutrino signal expected from a galactic supernova explosion from the
 cosmic-ray related backgrounds.
 This extra-capability of the MiniBooNE detector causes no disruption to its main
 accelerator-based $\nu_{\mu}\rightarrow \nu_e$ search.
%
%
\section*{Acknowledgments}
I am particularly indebted to Janet Conrad, Matthew Sharp, John Beacom and Joseph
 Formaggio, for the suggestions and help provided in preparing these proceedings,
 and to the Organizing Committee for the stimulating atmosphere of the 37th Rencontres
 de Moriond on Electroweak Interactions and Unified Theories.

\end{document}